\documentclass{article}

\usepackage[final]{neurips_2023_ml4ps}

\usepackage[utf8]{inputenc} 
\usepackage[T1]{fontenc}    
\usepackage{hyperref}       
\usepackage{url}            
\usepackage{booktabs}       
\usepackage{amsfonts}       
\usepackage{nicefrac}       
\usepackage{microtype}      
\usepackage{xcolor}         
\usepackage{amsmath}
\usepackage{graphicx}

\title{$\rho$-Diffusion: A diffusion-based density estimation framework for computational physics}

\author{%
  Maxwell X. Cai \\
  Intel Corporation\\
  \texttt{maxwell.cai@intel.com} \\
  \And 
  Kin Long Kelvin Lee \\
  Intel Corporation \\
  \texttt{kin.long.kelvin.lee@intel.com}
}

\begin{document}

\maketitle

\begin{abstract}
In physics, density $\rho(\cdot)$ is a fundamentally important scalar function to model, since it describes a scalar field or a probability density function that governs a physical process. Modeling $\rho(\cdot)$ typically scales poorly with parameter space, however, and quickly becomes prohibitively difficult and computationally expensive. One promising avenue to bypass this is to leverage the capabilities of denoising diffusion models often used in high-fidelity image generation to parameterize $\rho(\cdot)$ from existing scientific data, from which new samples can be trivially sampled from. In this paper, we propose \textsc{$\rho$-Diffusion}, an implementation of denoising diffusion probabilistic models for multidimensional density estimation in physics, which is currently in active development and, from our results, performs well on physically motivated 2D and 3D density functions. Moreover, we propose a novel hashing technique that allows \textsc{$\rho$-Diffusion} to be conditioned by arbitrary amounts of physical parameters of interest. 
\end{abstract}

\section{Introduction}

High-quality scientific data---the foundation of all scientific discovery---is difficult and expensive to accumulate. The time and energy cost to design, implement, and execute state-of-the-art experiments (\textit{in silico} and \textit{ex silico}) necessary for many domains remains barely surmountable for all but select problems, further gated behind access to the required resources such as equipment, human expertise, compute, and so on. A significant amount of scientific research could be enabled and/or accelerated if the availability of critical data was improved.

At risk of being overly reductive, a significant amount of scientific research pertains to modeling probability densities, $\rho(\cdot)$; from a probabilistic perspective, by estimating or learning distributions of parameters that govern some physical process, new data and inferences can in principle be obtained through posterior sampling. Traditional density estimators, such as histograms, kernel density estimators (KDEs), and Markov chain Monte Carlo (MCMC) only perform well in low dimensions. But the recent renaissance of generative modeling with deep neural networks has spawned a plethora of approaches ranging from variational inference \citep{2013arXiv1312.6114K} to adversarial learning \citep{2014arXiv1406.2661G} to normalizing flows \citep{2015arXiv150505770J,2018arXiv180607366C} to diffusion models \citep{pmlr-v37-sohl-dickstein15,2020arXiv200611239H}. While most of these methods have demonstrated in various capacities both the potential and ability to be applied to the physical sciences (see e.g., \cite{2015arXiv150203509G,2016arXiv160508803D,2017arXiv170507057P,doi:10.1073/pnas.2101344118,2018arXiv181001367G}), much of the research focus has remained in select applications particularly in the 2D image domain or toy datasets. Applications of denoising diffusion probabilistic models (DDPM) in physics in particular has seen success in several areas (e.g., \cite{2022PNAS..11903656W,2022MNRAS.511.1808S,2023JCoPh.47811972S,2023arXiv230602929L}). 

In this paper, we propose a DDPM-based density modeling framework that provides scientific domain-centered abstractions for conditional data generation, with the primary goal being to lower the barrier for adopting generative practices to scientific workflows. As a proof-of-concept, we demonstrate cross-domain adaptability with data generation in astrophysics and chemistry for 2D/3D data.

\section{Methods}
\paragraph{Denosing Diffusion Probabilistic Models} Inspired by nonequilibrium thermodynamics, \citep{pmlr-v37-sohl-dickstein15} showed that the gradual and systematic noise injection to input data can be considered a Markov process. Given $\rho^{(t)}(\cdot)$ as the density function of interest at diffusion timestep $t$, by using a tractable Gaussian $\mathcal{N}(\rho^{(t)}; \sqrt{1 - \beta^{(t)}} \rho^{(t)}, \beta_t \mathbf{I})$ as the diffusion kernel, the diffusion trajectory is deterministic, and with a sufficiently large number of steps $T$, $\rho^{(T)} \rightarrow \mathcal{N}(\mathbf{0}, \mathbf{I})$. Conversely, the data generation process is elegantly translated into learning a reverse trajectory that starts from $\mathcal{N}(\rho^{(T)}; \mathbf{0}, \mathbf{I})$ and ends at $\rho^{(0)}(\cdot)$. Notably, if the schedule steps size $\delta \beta := \beta_{t} - \beta_{t-1}$ is sufficiently small, then the reverse kernel will have the same functional form as the diffusion kernel, i.e., Gaussian. Therefore, the neural network only has to predict the correct time-dependent mean of the reverse Gaussian kernel (the variance of which can be computed directly from $\beta^{(t)}$). Eventually, $\rho^{(0)}$ can be sampled from pure Gaussian noise through a Langevin-like reverse process \citep{10.5555/3104482.3104568}. New data that is not part of the original training data can be obtained by interpolating the training data \citep{2020arXiv200611239H}. For details, we refer the readers to the original papers \citep{pmlr-v37-sohl-dickstein15, 2020arXiv200611239H}.

\paragraph{Architecture} A neural network that learns $\mu_{\theta}(\rho^{(t)}, t)$ must be able to accept $n$-dimensional $\rho^{(t)}(\cdot)$ at an arbitrary timestep $t$. We follow the practice in e.g., \cite{2020arXiv200611239H,2021arXiv210505233D} to use a U-Net \citep{2015arXiv150504597R} for the parameterization, which is enhanced with residual \citep{2015arXiv151203385H} building blocks, multi-head attention \citep{2017arXiv170603762V}, and a Transformer-style sinusoidal position embedding to allow parameters to be shared across time.
Additionally, thanks to the close form of $q(\rho^{(t)}|\rho^{(0)}) = \mathcal{N} \left( \rho^{(t)}; \sqrt{\Bar{\alpha}^{(t)}} \rho^{(0)}, (1-\Bar{\alpha}^{(t)})\mathbf{I} \right)$ where $\Bar{\alpha}^{(t)} := \prod_{s=1}^{t} (1-\beta^{(s)}$), it is possible to obtain a noised sample $\rho^{(t)}$ directly from $\rho^{(0)}$ \citep{2020arXiv200611239H}. Therefore, instead of looping over $t$, training can be accelerated in a (distributed) data-parallel fashion. 

\paragraph{Implementation} We are aware of the existence of many open-source DDPM implementations. While most existing implementations are designed for 2D images, a few (e.g., \textsc{Diffusers} from HuggingFace and \textsc{GuidedDiffusion} from OpenAI) have extended the API to 1D and 3D data. However, most of these implementations are optimized for tasks like image generation or text-to-image translation, and the underlying software architectures do not necessarily provide the flexibility to accommodate the domain-specific requirements and properties of physical data. As such, we eventually opted for a customized, physics-centric implementation from the stretch, which allows us to integrate physics-motivated models and inductive biases into the framework in the subsequent development. We generalized the DDPM backbone of \textsc{$\rho$-Diffusion} to $n$-dimensional, with a current limitation of $n = \{1, 2, 3\}$, which corresponds to sequential data, 2D PDFs/images, and 3D PDF/volumetic data, respectively. The model is implemented with \textsc{PyTorch}, Intel\textsuperscript{\textregistered} Extension for PyTorch (IPEX) and Intel\textsuperscript{\textregistered} oneAPI Deep Neural Network Library (oneDNN). The model training and inference were performed using Intel\textsuperscript{\textregistered} Data Center GPU Max Series.

\paragraph{Conditioned data generation with physical parameters} In physics, generative models are particularly useful if the data generation process can be conditioned directly by the physical parameters of interest. In the DDPM case, it boils down to learning $p \left( \rho^{(t)} | \rho^{(t-1)}, c \right)$, where $c$ is the condition. In the traditional class-conditioned DDPM image generation approach, the class labels are typically categorical, which are then passed to an embedding layer to calculate the embedding vector. Such an approach is inadequate to handle the scenario where $\rho(\cdot)$ is a continuous and multivariate function. Instead, we allow the input variables of $\rho(\cdot)$ to be expressed in a \textsc{Python} \texttt{dict} structure and subsequently use the \texttt{SHA512} algorithm to hash the parameter \texttt{dict}. The hexadecimal digest is then converted into a float vector, which serves as the embedding of \texttt{params}. The uniqueness of the embedding is guaranteed by the \texttt{SHA512} algorithm, which is deterministic and has a hash collision rate of $1/2^{512}$, and the length of the embedding is fixed, independent of the number of parameters given. This makes it possible for \textsc{$\rho$-Diffusion} to be conditioned by an arbitrary amount of physical parameters of interest. 

This hash-based embedding approach is computationally efficient, and by virtue of being non-parametric, does not require domain adaptation and can be used for arbitrary formats. Its main weakness however is that it is not possible to interpolate the model directly in the physical parameter space: while the avalanche effect is a desirable property for cryptography, minuscule changes in the original parameters will result in a completely different embedding. In the case of \textsc{$\rho$-Diffusion}, ideally the embedding function is not only differentiable, but also maps perturbations from physical parameters to the embedding space in a physically meaningful capacity. A possible improvement in future work---inspired by \textsc{CLIP}\citep{2021arXiv210300020R}---is to use train a language model jointly with the DDPM, along with the physical parameters as the guidance to condition the $\rho(\cdot)$ sampling. 


\section{Experiments}
In the following experiments, we demonstrate that \textsc{$\rho$-Diffusion} works on both traditional image data and real physical data (with proper normalization, if necessary). 

So far, we have shown success in applying our workflow to 2D and 3D data: the combination of UNet employing either 2D and 3D convolutions with Langevin sampling can produce conditional densities with high fidelity and accuracy. However, we see the approach break down for the 1D case of pure rotational spectra, which correspond to observable transitions between quantized rotational energy levels of molecules \citep{townesMicrowaveSpectroscopy1975}. This demonstrates a number of complexities not captured: notably, regions of high ``density'' or signal are significantly more sparse in high-resolution molecular spectra than the other data covered here and likely require architectures that recover long-range correlations.

\begin{figure}
    \centering
    \includegraphics[scale=0.12]{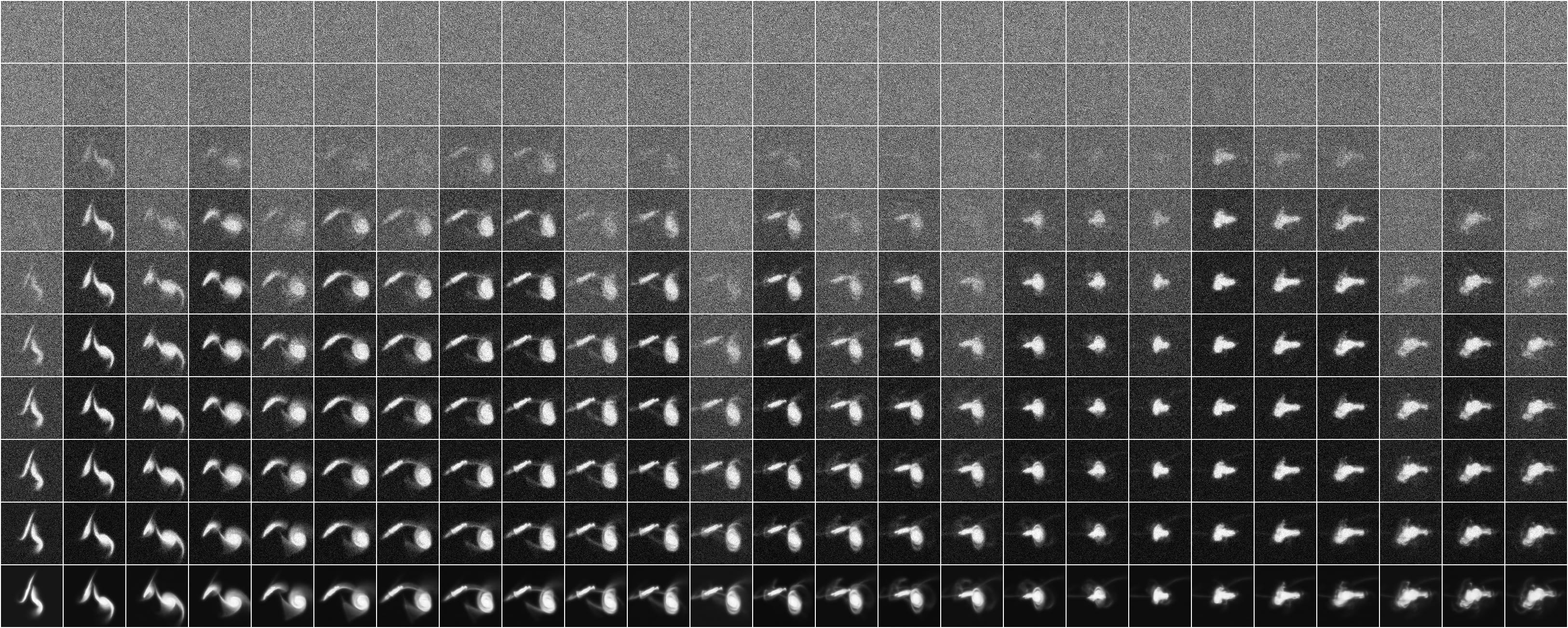}
    \caption{Generated galaxy merger image samples (128x128 px). From top to down: the denoising process at $t=[900, 800, ...,100, 0]$, respectively; From left to right: the merging process from initial approaching to tidal disruption to collision, where each column corresponds to a snapshot $5 \times 10^6$ years apart from the adjacent columns in physical timescale.}
    \label{fig:2d_samples}
\end{figure}

\paragraph{2D Data: Images of galaxy merger simulations} Galaxy mergers are fundamentally important processes driven through the history of galactic evolution \citep{1977egsp.conf..401T, 1992ARA&A..30..705B}. In this process, the two colliding galaxies are tidally disrupted, creating a time-evolving morphology. It is computationally expensive to simulate galaxy mergers. In this experiment, we use a Barnes-Hut tree \citep{1986Natur.324..446B} code \textsc{Bonsai} \citep{2012JCoPh.231.2825B} to simulate the dynamical process with a grid of parameters and subsequently generate 2D images from the visualization simulation snapshots. The 2D images are then used to train  \textsc{$\rho$-Diffusion}, in which the physical parameters used to run \textsc{Bonsai} are used to condition the sampling. Figure~\ref{fig:2d_samples} shows that by varying the time parameter while keeping the others fixed, \textsc{$\rho$-Diffusion} can generate the merger process directly, thus bypassing the need to carry out expensive simulations.

\begin{figure}
    \centering
    \includegraphics[scale=0.25]{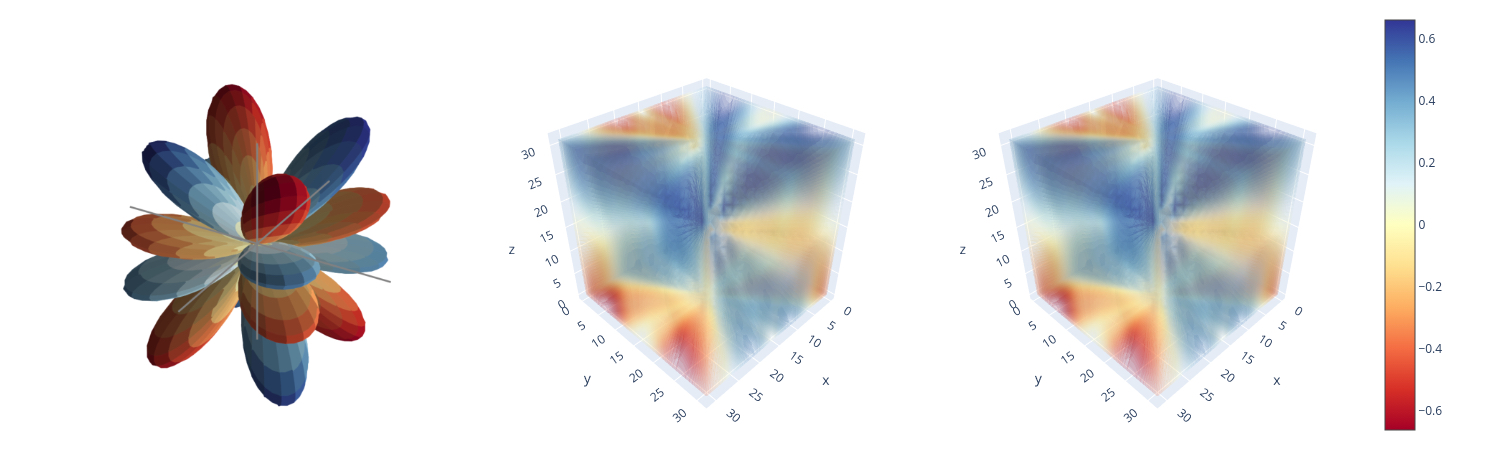}
    \caption{Left: A surface plot of $Y_{5}^{2}(\theta, \phi)$; Center: the voxelized ground truth volume; Right: the generated volume. The real part is color-coded, and all three subplots share the same colorbar. The generated volume is visually indistinguishable from the ground truth, and has a Wasserstein distance of ${\sim}0.035$ from the ground truth. }
    \label{fig:3d_samples}
\end{figure}

\paragraph{3D data: Volumes of spherical harmonics} Spherical harmonics are a mathematical basis with diverse applications, ranging from molecule spectroscopy \citep{townesMicrowaveSpectroscopy1975} and quantum chemistry \citep{schlegelTransformationCartesianPure1995} to astronomy \citep{brettMethodsSphericalHarmonic1988}. In the context of density estimation, spherical harmonics are an attractive source of data to generate and test arbitrarily complex and structured two and three-dimensional densities; in spherical coordinates, each spherical harmonic $Y$ is defined or ``conditioned'' by two parameters, the order $m$ and degree $l$:
\begin{equation}
    Y^m_l(\theta, \phi) = \sqrt{\frac{2n + 1}{4\pi} \frac{(n-m)!}{(n + m)!} e^{im\theta} P^m_l \left(\cos(\phi) \right)}
    \label{eq:spherical-harmonics}
\end{equation}
with associated Legendre polynomial $P^m_l$, and azimuthal and polar coordinates $\theta$ and $\phi$. Larger values of $l$ and $m$ progressively introduce symmetric nodal structure, which may prove increasingly difficult for models to reproduce. To construct these basis functions as 3D objects, we rely on their implementation in \textsc{NumPy} \citep{harris2020array} and \textsc{SciPy} \citep{2020SciPy-NMeth} and compute them on linearly spaced cartesian grids, with the value at each point corresponding to the real part of $Y^m_l(x,y,z \rightarrow \theta, \phi)$. We compute the spherical harmonics for different combinations of $(m, l)$ and voxelize them to volumes of $32^3$. Figure~\ref{fig:3d_samples} shows a non-cherrypicked generated sample of $Y_{5}^{2}(\theta, \phi)$.

\section{Discussions and Conclusions}
DDPM models are known for their capability of high-quality image density synthesis. In this project, we extend this capability to physics-based density estimation tasks. Theoretically, the DDPM framework places no limits on the types of density functions. In practice, however, we observe several difficulties when training DDPM to learn $\rho(\cdot)$ on physical data due to their unique properties:
\begin{itemize}
    \item Large temporal ranges: Sequential data, especially time series of a dynamical system, may exhibit long-range dependencies. Despite the enhancements on the U-Net with attention blocks and residual networks, the model may not be informed about the global structure of the sequence due to the small convolutional kernel sizes. 
    \item Large spatial ranges: in many fields in physics, especially in astrophysics, the range of the data is common to cross multiple orders of magnitudes. In the DDPM framework, the use of $ \mathcal{N}(\rho^{(t)}; \sqrt{1 - \beta_t} \mathbf{x}^{(t-1)}, \beta_t \mathbf{I}) $ as a diffusion kernel has essentially required the data to be scaled to a small internal, ideally $[-1,1]$. However, squeezing raw physical data into such a range will sacrifice the model's resolution on small details. 
    \item Lack of inductive biases: Although diffusion models are inspired by nonequilibrium thermodynamics, the DDPM framework does not necessarily place physical constraints on an architectural level. This may cause slower convergence and limitations on the model's capability to understand the physical laws behind the data.
    \item Expensive sampling process: Unlike other generative models, sampling DDPM is an iterative process that typically requires ${\sim}10^3$ steps of model inference. The nonlinear behavior in physical data may require even more steps due to the fluctuating gradients. We expect this issue to be alleviated with improved sampling algorithms (e.g., DDIM sampling\cite{2020arXiv201002502S}).
\end{itemize}


Despite the challenges, we demonstrate that under certain conditions, DDPM-based models are capable of capturing the fine details of some density functions in physics. As a proof-of-concept, we propose \textsc{$\rho$-Diffusion} as a framework to handle physical data of different dimensionalities with a unified application programming interface (API), and provide a workflow for training and inference. The framework is currently in active development\footnote{Source code: \href{https://github.com/intel/rho-diffusion}{\tt https://github.com/intel/rho-diffusion}}  and contributions are welcome.




\medskip

{
\small
\bibliographystyle{unsrt}
\bibliography{refs}
}


\end{document}